\documentstyle[aps,prd,eqsecnum,amssymb,amsfonts,verbatim]{revtex}
\renewcommand{\H}{{\cal H}}
\newcommand{\bs}{\bbox}
\newcommand{\p}{\hat{p}}
\newcommand{\<}{\langle}
\newcommand{\be}{\begin{equation}}
\renewcommand{\>}{\rangle}
\newcommand{\s}{\sigma}
\newcommand{\sm}{{\mathsf{s}}}

\begin{document}
\title{Relativistic Partial Wave Analysis Using the Velocity
Basis of the Poincar\'e Group} 
\author{A.~Bohm}
\address{The University of Texas at Austin\\
Austin, Texas 78712\\ 
bohm@physics.utexas.edu}
\author{H.~Kaldass}
\address{The University of Texas at Austin\\
Austin, Texas 78712\\
hani@physics.utexas.edu}
\date{\today}
\maketitle
\begin{abstract}
The velocity basis of the Poincar\'e group is used in the
direct product space of two irreducible unitary representations
of the Poincar\'e group. The velocity basis with total
angular momentum $j$ will be used for the definition of relativistic
Gamow vectors.
\end{abstract}
\maketitle

\section{Introduction}

Resonances are obtained in the scattering of two (or more) elementary
particles, and quasistationary states decay into a two (or many) particle 
system with masses $m_{i}$ and spins $s_{i}$, $i=1,\,2\cdots$. 
Relativistic resonances and 
decaying states are therefore described in the direct product space of two
irreducible representation spaces of the Poincar\'e group
${\cal H}={\cal H}_{1}(m_{1},s_{1})\otimes{\cal H}_{2}(m_{2},s_{2})$.
Non-relativistic resonances and decaying states
have been described by Gamow vectors \cite{our}. 
Gamow vectors are characterized by a 
value of angular momentum $j$ in the center-of-mass frame and by a complex
energy $z_{R}=\left(E_{R}-i\frac{\Gamma}{2}\right)$, representing
resonance energy $E_{R}$ and lifetime $\frac{\hbar}{\Gamma}$. They
are generalized eigenvectors in a Rigged Hilbert Space
$\Phi\subset\H\subset\Phi^{\times}$ of the self-adjoint Hamiltonian
$H$ with complex eigenvalue $z_{R}$ \cite{our}.
Relativistic resonances and unstable particles
are characterized by their spin (total angular momentum in the
center-of-mass frame of the decay products) and the value $\sm=\sm_{R}
\equiv \left(M_{R}-i\frac{\Gamma}{2}\right)^{2}$ of the invariant mass
squared $\sm=(p_{1}+p_{2})^{2}=(E^{2}-\bs{p}^{2})$
where $M_{R}$ is the resonance mass and $\frac{\hbar}{\Gamma_{R}}$
is its lifetime. We want to find relativistic Gamow vectors which
are generalized eigenvectors of the total mass operator
$M^{2}=P_{\mu}P^{\mu}=(P_{1\mu}+P_{2\mu})(P_{1}^{\mu}+P_{2}^{\mu})$ 
with complex eigenvalue $\sm_{R}$
and with spin $j$. These must be obtained from the direct
product space ${\cal H}_{1}(m_{1},s_{1})\otimes{\cal H}_{2}(m_{2},s_{2})$.

Eigenspaces of $M^{2}$ with real values of invariant mass $\sm$
and total angular momentum $j$ are obtained by the relativistic partial
wave analysis \cite{aW60,hJ62,aM62} 
using the Wigner basis, i.e., using momentum eigenvectors
$|\bs{p}_{i},s_{3i}(m_{i},s_{i})\>$ in the spaces
${\cal H}_{i}$ and eigenvectors $|\bs{p},j_{3}(\sm,j)\>$ 
of $P_{\mu}=P_{1\mu}+P_{2\mu}$ in the direct 
product space $\cal H$. 

In distinction to the non-relativistic
case, in the relativistic case Lorentz transformations intermingle energy 
and momenta. If one wants to make an analytic continuation
of $\sm$ from the values $(m_{1}+m_{2})^{2}\leq \sm<\infty$
to the complex values $\sm_{R}$ (of the pole position in the second 
sheet of the relativistic $S$-matrix $S_{j}(\sm)$) this will also
lead to complex momenta. To restrict the unwieldy 
set of complex momentum representations \cite{cm} we want to construct
complex mass representations of the Poincar\'e group $\cal P$ whose
momenta  are ``minimally complex''
in the sense that though $p_{\mu}$ and $m$ are complex, the 
$4$-velocities $\p_{\mu}\equiv \frac{p_{\mu}}{m}$ remain real.
This can be carried out because, as explained 
in section $2$, the $4$-velocity eigenvectors
$|\bs{\p},j_{3}(\sm,j)\>$ 
provide as valid basis vectors for the representation space
of $\cal P$ as the usual momentum eigenvectors. 
Moreover, they are more useful for physical reasoning than the 
momenta eigenvectors, because the $4$-velocities
seem to fulfill to rather good approximation ``velocity super-selection
rules'' which the momenta do not \cite{velocityvectors}.
Therefore we will use the velocity basis $|\bs{\p}_{i},s_{3i}(m_{i},s_{i})\>$
for the relativistic partial wave analysis and obtain the Clebsch-Gordan
coefficients of the Poincar\'e group for the velocity basis. This is done 
in section $3$ for $s_{1}=s_{2}=0$, 
which applies to the case of $\pi^{+}\pi^{-}$
in the final state.
This gives the velocity eigenvectors $|\bs{\p},j_{3} (\sm,j)\>$ 
of the direct product space ${\cal H}=\sum_{j=0}^{\infty}
\int_{(m_{1}+m_{2})^{2}}^{\infty}d\mu(\sm){\cal H}(\sm,j)$
from which we obtain the 
four-velocity scattering states $|\bs{\p},j_{3} (\sm,j)^{\pm}\>$
using the Lippmann-Schwinger equation as e.g., done in
\cite{sW95}. The relativistic Gamow vectors
$|\bs{\p},j_{3}(\sm_{R},j)^{\pm}\>$
will be obtained in a subsequent paper from the scattering states
by analytic continuation. In the Appendix, we derive the Clebsch-Gordan
coefficients for the velocity basis of $\cal P$ for the general case.


\section{Velocity Basis of the Poincar\'e Group}

We denote the ten generators of the unitary representation
${\cal U} (a, \Lambda)$ of $(a, \Lambda)\in \cal{P}$, by
\begin{equation}
\label{generators}
P^{\mu},\, J^{\mu\nu}{\hspace{1cm}} \mu,\nu\,=\, 0,1,2,3\, .
\end{equation}
The standard choice of the invariant operators and of a complete 
set of commuting observables (c.s.c.o.) is
\begin{eqnarray}
\nonumber
M^{2}=P_{\mu}P^{\mu}\,&,&\quad W=-w_{\mu}\,w^{\mu},\\
\label{csco}
P_{i}\,(i=1,2,3)\,&,&\quad S_{3}=M^{-1}
{\cal U}(L(p))\,w_{3}\,{\cal U}^{-1}(L(p))\, ,
\end{eqnarray}
here 
\begin{equation}
\label{w}
w_{\mu}=\frac{1}{2}\,\epsilon_{\mu \nu \rho \sigma}\, P^{\nu}
J^{\rho \sigma}\, ,
\end{equation}
$M^{-1}$ is the inverse square root of the positive definite operator
$P^{\mu}P_{\mu}$,~and ${\cal U}(L(p))$ is the representation of the boost that
depends upon the parameters $p_{\mu}\,(\mu=0,1,2,3)$, which are
the eigenvalues of the operators 
$P_{\mu}$. Only three of these parameters are independent 
in an irreducible representation, because of 
the relation $m^{2}=p_{\mu}p^{\mu}$. The standard boost 
(``rotation free'') matrix $L^{\mu}_{.\, \nu}(p)$ is given by
\begin{equation}
\label{standardboost}
L^{\mu}_{.\,\nu}(p)=
\bordermatrix{     &\nu=0           &\nu=n \cr
              \mu=0&\frac{p^{0}}{m} &-\frac{p_{n}}{m} \cr
               \mu=m&\frac{p^{m}}{m} &\delta^{m}_{n}-
                          \displaystyle{\frac{\frac{p^{m}}{m}\, 
\frac{p_{n}}{m}}
					{1+\frac{p^{0}}{m}}}\cr }\, .
\end{equation}   
Note that $p_{\mu}=\eta_{\mu\,\nu}p^{\nu}$ and we use the metric 
$\eta_{\mu\,\nu}=\scriptstyle{\left(
\begin{array}{cccc}
1 &   &    & 0 \\
  &-1 &    &   \\
  &   & -1 &   \\
0 &   &    & -1 
\end{array}\right)}\, $
\footnote{Some of the references we use here have different convention,
e.g., $\eta_{\mu\, \nu}\rightarrow -\eta_{\mu\, \nu}$ \cite{sW95}, 
and $L^{-1}\rightarrow L(p)$ \cite{hJ62}.}.
It has the property that
\begin{equation}
\label{standardmomentum}
L^{-1}(p)^{\mu}_{.\,\nu}p^{\nu}=
\left(\begin{array}{c}
m\\
0\\
0\\
0
\end{array}\right)\, .
\end{equation}
One feature shown in (\ref{standardboost}) 
which we want to make use of, is that the
boost $L^{\mu}_{.\,\nu}(p)$ does not depend upon $p$ but only upon the
4-velocity $\frac{p}{m}\equiv \p$. The complete basis system in the irreducible
representation space ${\cal H}(m^{2},j)$ which consists of eigenvectors
of the c.s.c.o.\ (\ref{csco}) is the Wigner basis usually denoted as
\begin{equation}
\label{wignerbasis}
|\bs{p},j_{3}(m,j)\rangle \, .
\end{equation}
It has the transformation property under the translation $(a,{\rm I})$ and the
Lorentz transformation $(0,\Lambda)$ :
\begin{mathletters}
\label{poincaretransformations}
\begin{equation}
\label{translation}
{\cal U}(a,{\rm I})|{\bs p},j_{3}\rangle=
e^{ip^{\mu}a_{\mu}}|{\bs p},j_{3}
\rangle
\end{equation}
\begin{equation}
\label{lorentztransformation}
{\cal U}(0,\Lambda)|{\bs p},\xi\rangle=\sum_{\xi^{'}}|
\bs{\Lambda p},
\xi^{'}\rangle D_{\xi^{'}\xi}({\cal R}(\Lambda,p))\, ,
\end{equation}
where $\cal{R}$ is the Wigner rotation
\begin{equation}
\label{wignerrotation}
{\cal R}(\Lambda,p)=L^{-1}(\Lambda p)\Lambda L(p)\, .
\end{equation}
The Wigner rotation depends upon the 
$10$ parameters of $\Lambda$ and upon the parameters 
$\p^{\mu}=\frac{p^{\mu}}{m}\, .$
In an UIR there are $3$ independent $\p^{\mu}$ and :
\begin{equation}
\label{boost}
|p,j_{3}\rangle={\cal U}(L(p))|{\bs p}={\bs 0}, 
j_{3}\rangle \, ,
\end{equation}
\end{mathletters}
where we have omitted the fixed values $m\, j$ as we shall often do 
in an UIR.
Every vector (of a dense subspace of physical states) of ${\cal H} (m,j)$
can be written according to Dirac's basis vector decomposition as
\begin{mathletters}
\label{diracbasisvectors}
\begin{equation}
\label{dirac}
\phi=\int \, d\mu\left(\bs{p}\right)\sum_{\xi}|\bs{p},\xi
\rangle 
\langle {\bs p},\xi\,|\,\phi\rangle \, ,
\end{equation}
where one has many arbitrary choices for the measure.
It is usually chosen to be given by
\begin{equation}
\label{measure}
d\mu\left(\bs{p}\right)=\rho\left(\bs{p}\right)d^{3}
\bs{p}\, ,
\end{equation}
where one can choose any (measurable) function $\rho$, in particular
a smooth function. The choice of $\rho$ is connected to the 
``normalization'' of the Dirac kets through :
\begin{equation}
\label{normalization}
\langle \xi^{'},\bs{p'}\,|\,\bs{p},\xi \rangle
=\frac{1}{\rho\left(\bs{p}\right)}\, \delta^{3}
(\bs{p}-{\bs p'})\, 
\delta_{\xi \xi'} \, .
\end{equation}
One convention\footnote{This is the convention of \cite{eW39,aW60,hJ62,aM62}, 
but not
of \cite{sW95}} for $\rho$
is the Lorentz invariant measure :
\begin{equation}
\label{invariantmeasure}
\rho\left(\bs{p}\right)=\frac{1}{2E(\bs{p})}
\, ,\quad\text{where }
E\left(\bs{p}\right)=\sqrt{m^{2}+{\bs{p}}^{\,2}}\, .
\end{equation}
\end{mathletters}

The mathematically precise form of the Dirac decomposition is
the Nuclear Spectral Theorem for the complete system of 
commuting (essentially self-adjoint) operators. It is the same as
(\ref{diracbasisvectors}), however with well
defined mathematical quantities. The state vectors $\phi$ 
in (\ref{dirac}) must be elements of a 
dense subspace $\Phi$ of the representation space $\H$ of an UIR : 
\begin{equation}
\phi \in \Phi \subset \H(m,j)\,;
\end{equation} 
and the basis vectors $|\bs{p},\xi\rangle \in \Phi^{\times}$
are elements of the space of antilinear functionals
on $\Phi$ which fulfill the condition :
\begin{mathletters}
\begin{equation}
\label{generalizedeigenvector}
\langle P_{i} \psi \,|\, \bs{p}, \xi \rangle=p_{i}\, 
\langle \psi \,|\,
\bs{p}, \xi
\rangle \quad\text{for every }\psi \in \Psi\, .
\end{equation}
This condition means the $|\bs{p},\xi\>$ are generalized eigenvectors 
of $P_{i}$, which is
also written as 
\begin{equation}
P_{i}^{\times}|\bs{p},\xi\>=p_{i}\,|\bs{p},\xi\>\, ,
\end{equation}
\end{mathletters}
where $P_{i}^{\times}$ is an extension of $P_{i}^{\dagger}(=P_{i})$;
and the ``component of $\phi$ along the basis vector
$|\bs{p},\xi\rangle$'', the 
$\langle \bs{p},\xi\,|\,\phi \rangle=\<\phi\,|\,\bs{p},\xi\>^{*}
$, are antilinear
continuous functionals $F(\phi)=\langle \bs{p},\xi|\phi\rangle^{*}$ 
on the space $\Phi$.

The space $\Phi$ is a dense nuclear subspace of $\cal{H}$ \cite{aB73}. 
(E.g., $\Phi$
could be chosen to be the subspace of
differentiable vectors of $\cal{H}$ equipped with a nuclear topology
defined by the countable number of norms~: 
$||\phi||_{p}=\sqrt{(\phi,(\Delta+1)^{p}\phi)}$, where
$\Delta=\sum_{\mu} P_{\mu}^{2}+\sum_{\mu \, \nu}\frac{1}{2}J^{2}_{\mu \, \nu}
$ is the Nelson operator \cite{eN59}. But it could also
be chosen as another dense nuclear subspace of $\cal{H}$.)
The three spaces form a Gel'fand triplet, or Rigged Hilbert Space
\begin{equation}
\label{RHS}
\Phi \, \subset \, {\cal H} \, \subset \, \Phi^{\times}
\end{equation}
and the bra-ket $< \, | \, >$ is an extension of the scalar product
$( \, , \, )$. The $\langle \bs{p}, \xi\,|\,\phi \rangle
=\<\phi\,|\,\bs{p}, \xi\>^{*}$ are the Wigner momentum 
wavefunctions.

The Wigner kets (\ref{wignerbasis}) 
are not the only basis system of ${\cal H}(m,j)$
that one can use to expand every vector $\phi \in \Phi$. For every 
different choice of c.s.c.o.\
in the enveloping algebra $\cal{E}({\cal P})$ (the algebra generated by
$P_{\mu}$, $J_{\mu \, \nu}$) one obtains a different system
of basis vectors; in this
way one can obtain e.g., Lorentz basis 
(eigenvectors of the Casimir operators of
$SO(3,1)_{J_{\mu \nu}}$ \cite{hJ62,aB73}), or 
the spinor basis (whose Fourier transforms are
the relativistic fields \cite{sW95}) etc. 
We want to choose still another basis system, 
which is similar to the Wigner basis except that it is 
a basis of eigenvectors of the $4$-velocity operator
$\hat{P}_{\mu}\equiv P_{\mu}M^{-1}$ rather than the momentum operator
$P_{\mu}$. 

With the 4-velocity operator, one defines the operators
\begin{equation}
\label{what}
\hat{w}_{\mu}=\frac{1}{2}\,\epsilon_{\mu \nu \rho \sigma}
\hat{P}^{\nu}J^{\rho  \sigma}=w_{\mu}M^{-1} \, ,
\end{equation}
and the spin tensor
$$
\Sigma_{\mu \, \nu}=\epsilon_{\mu \nu \rho \sigma}\hat{P}^{\rho}
\hat{w}^{\sigma}\,.
$$
The c.s.c.o.\ is then given by
\begin{equation}
\label{cscohat}
\hat{P}_{m},\quad S_{3},\quad \hat{W}=-\hat{w}_{\mu} \hat{w}^{\mu}
=\frac{1}{2}\Sigma_{\mu \nu}\Sigma^{\mu \nu},\quad M^{2}\, ,
\end{equation}
and we denote its generalized eigenvectors by
\begin{equation}
\label{wignerbasishat}
|\bs{\p},j_{3};\sm=m^{2},j\>\, ,
\end{equation}
where $\p_{\mu}=\frac{p_{\mu}}{m}$ are the eigenvalues of $\hat{P}_{\mu}$.

The basis vector expansion for every $\phi \in \Phi$ 
with respect to the basis system (\ref{wignerbasishat}) is given by
\begin{mathletters}
\label{diracbasisvectorhat}
\begin{equation}
\label{dirachat}
\phi=\sum_{j_{3}}\int \, \frac{d^{3}\p}{2 \p^{0}}
\,|\bs{\p}, j_{3}\rangle \langle j_{3}, \bs{\p} 
\,|\, \phi \rangle \, , 
\end{equation}
where we have chosen the invariant measure
\begin{eqnarray}
\label{measurehat}
d\mu(\bs{\p}) &=& \frac{d^{3}\p}{2\p^{0}}
               = {\frac{1}{m^{2}}} \, {\frac{d^{3}p}{2 E(\bs{p})}}\\             \nonumber  \p^{0} &=& \sqrt{1+\bs{\p}^{2}} \, .
\end{eqnarray}
As a consequence of (\ref{measurehat}), 
the $\delta$-function normalization of these velocity-basis vectors is
\begin{eqnarray}
\nonumber
\langle \xi , \bs{\p}\,|\,\bs{\p'}, \xi' \rangle
       &=& 2 \p^{0}  \delta^{3}(\bs{\p}-\bs{\p'})
                              \, \delta_{\xi \xi'}\\
\label{normalizationhat} 
 &=& 2 p^{0} m^{2} \delta^{3}
(\bs{p}-\bs{p'})\, 
\delta_{\xi \xi'} \, .
\end{eqnarray}
\end{mathletters}

Mathematically, every c.s.c.o.\ is equally valid. 
But, for a given physical problem one c.s.c.o.\ may be more useful than 
another. For instance a c.s.c.o.\ that contains physically 
distinguished observables 
(e.g., observables whose eigenstates happen to appear predominantly in nature)
 is more useful for calculations in physics than the c.s.c.o.\ whose
eigenvectors are very different from physical eigenstates.
Two different c.s.c.o.'s lead to different basis systems, whose vectors can be 
expanded with respect to each other. But this expansion is usually very
complicated and intractable, for which reason the choice of the physically 
right c.s.c.o.\ is very important for each particular physical problem. This 
is the reason for which the Lorentz basis of the 
Poincar\'e group is pretty useless for physics, because
the Casimir operators of $SO(3,1)$ are not important observables as compared 
to the momentum. However, the two c.s.c.o.\ (\ref{csco}) and (\ref{cscohat})
are not even different in an irreducible representation of ${\cal P}$, 
since its operators differ only by a factor of 
the operator $M$, which is an invariant. The basis systems (\ref{wignerbasis}) 
and (\ref{wignerbasishat}) are therefore the same, 
i.e., their values differ by a normalization-phase factor $N(p,j_{3})$
\begin{equation}
\label{basisrelation}
|\,\bs{\p}, j_{3} \, (m,j)\rangle
=|\,{\bs p},j_{3} \, (m,j) \rangle \, N(p, j_{3}). 
\end{equation}
The Poincar\'e transformations (\ref{poincaretransformations}) 
act on the basis vectors (\ref{basisrelation}) in the following way
\begin{mathletters}
\label{poicaretranformationshat}
\begin{eqnarray}
&{\cal U}(a,{\rm I})|\bs{\p},j_{3}\rangle=
e^{im\p^{\mu}a_{\mu}}|\bs{\p}, j_{3}\rangle \label{translationhat}\\
&{\cal U}(L(\p))|\bs{\p}=\bs{0},j_{3}
\rangle=|\bs{\p},j_{3}\rangle \, .
\label{boosthat}
\end{eqnarray}
\end{mathletters}

The distinction between the basis vectors $|\,\bs{p}, \xi \rangle$
and $|\,\bs{\p}, \xi \rangle$ becomes important if one does not have an
unitary irreducible representation of $\cal{P}$ but a representation with
many different values for $(m^{2},j)$, e.g., ${\cal H}=\sum_{m^{2},j}
\oplus {\cal H}(m,j)$. Then one has besides the observables 
(\ref{generators}), additional observables 
$X_{\alpha}$ (generators of an intrinsic symmetry group
or a spectrum generating group) and an additional system of 
commuting observables :
\begin{equation}
\label{commutingobservables}
B=B_{1},B_{2}, \cdots , B_{N}
\end{equation}
whose eigenvalues, $b=(b_{1},b_{2}, \cdots ,b_{N})$,
characterize the elementary
particles described by ${\cal H}(m,j)={\cal H}^{b}(m,j)$
\footnote{The quantum numbers $b$ are called the particle species
numbers in \cite{sW95}.}.
In order that (\ref{csco}) and (\ref{commutingobservables}) combine
into a c.s.c.o., the operators $B$ have to commute with
$M^{2}, P_{\mu}, W \text{ and }S_{3}$. 
If also the other observables $X_{\alpha}$,
which change the particle species number $b$, commute with
$M^{2}, P_{\mu}, W \text{ and }S_{3}$, then the combination of 
(\ref{csco}) and (\ref{commutingobservables}) gives a useful
c.s.c.o. However, if the $X_{\alpha}$ do not commute with $M^{2}$
(i.e., the particle species number changing operators $X_{\alpha}$
transform also from one mass eigenstate to another mass eigenstate changing 
also the mass $m_{b}$ into $m_{b'}$) then the $X_{\alpha}$ will also
not commute with $P_{\mu}$, $\left[ X_{\alpha}, P_{\mu}\right]
\neq 0$. In this case, it may still happen \cite{velocityvectors} that a 
``velocity superselection rule'' holds :
\begin{equation}
\label{velocitysuperselectionrule}
\left[ X_{\alpha}, \hat{P}_{\mu} \right] =0\quad (\text{or at least }
\left[ X_{\alpha},\hat{P}_{\mu} \right]\approx 0)\, . 
\end{equation}
Then combination of (\ref{commutingobservables}) with 
(\ref{cscohat}), i.e., the
\begin{equation}
\label{cscohat2}
\hat{P}_{i},\, \hat{w}_{3},\, \hat{W},\, M^{2},\, B_{1},\cdots,\,B_{N}
\end{equation}
will form a useful c.s.c.o., but the combination of 
(\ref{csco}) with (\ref{commutingobservables}) will not.
The generalized eigenvectors of (\ref{cscohat2}),
$|\bs{\p},\xi,b,m,j
 \rangle $, will then be a much more useful basis system for every
$\phi \in \Phi \subset {\cal H}=\sum \oplus {\cal H}^{b}(m,j)$ than the
corresponding momentum eigenvectors.
Using the eigenvectors of (\ref{cscohat2}), we have the Dirac basis vector 
expansion :
\begin{equation}
\label{dirachat2}
\phi=\sum_{m,b}\sum_{j,\xi}\int \frac{d^{3}\p}{2\, \p^{0}}
|\,\bs{\p},\xi,b,m,j\rangle 
\langle j,m,b,\xi, \bs{\p}\,|\,
\phi \rangle \text{ for every } \phi \in \Phi\, .
\end{equation}
The  momentum eigenvectors $|\bs{p},\xi,b \ldots \rangle$
may either not exist (if $\left[ B, P_{\mu} \right] \not= 0$),
or if they do exist, they are not useful because the $X_{\alpha}$ 
change the value of $p$, which then becomes a function of $b$, 
$p=p_{b}$. 
As a consequence, quantities like form factors depend upon
$b$ through $p$. In contrast, using the velocity eigenvectors
$|\bs{\p},\xi,b,\cdots\rangle$ under the assumption
(\ref{velocitysuperselectionrule}) will lead to form factors with
universal (independent of $b$) dependence upon the four-velocity.
This was the original motivation for the introduction of 
the velocity-basis vectors $|\bs{\p},\xi,b,\cdots\rangle$
\cite{velocityvectors}.

The subject of the present work is the description of relativistic
decaying states by representations of the Poincar\'e group,
combining Wigner's idea \cite{eW39} of the description of
stable relativistic particles by an UIR of $\cal P$, with Gamow's 
idea of describing decaying particles by eigenvectors with
complex energy. Therefore, we need in the rest frame basis vectors 
with complex energy, i.e., the $m$ (and the $\sm=m^{2}$) in 
(\ref{wignerbasis}) or in (\ref{wignerbasishat})
has to be continued to complex values e.g., to $\sm=(M_{R}-i\Gamma/2)^{2}$. 
This will result in a continuation of the momenta $p_{\mu}$ to complex
values as well and can lead to an enormous complication of the 
Poincar\'e group representations (see e.g., \cite{cm}).
We want to do this analytic continuation in the invariant mass $\sm$
such that the $p_{\mu}$ are continued to complex values in such 
a way that the $\p_{\mu}=\frac{p_{\mu}}{\sqrt{\sm}}$ remain real. 
Then, we obtain a smaller class of complex
mass representations of $\cal P$ which are as similar in property
as possible to Wigner's UIR $(m,j)$. These are the minimally
complex-mass representations which we shall denote
by $(\sm,j)$.

For this minimal analytic continuation to be possible, it must be 
compatible with the boost (\ref{boost}) and (\ref{boosthat}). 
The crucial observation is that the boosts $L(p)$ are in fact, according to 
(\ref{standardboost}) only
functions of $\p_{\mu}=\frac{p_{\mu}}{\sqrt{\sm}}\,$; $L(p)=L(\p)$.
As a consequence, the operators representing the boost
${\cal U}(L(p))={\cal U}(L(\p))$ are functions of the real parameters
$\p$ and not of complex parameters $p$. 
This means they are the same operator functions in all the
subspaces of the
direct sum $\sum_{m_{b},j}\oplus{\cal H}(m_{b},j)$ and of the 
continuous direct sum  
\begin{equation}
\label{directsum}
\sum_{j,n}\int_{m_{0}^{2}}^{m_{1}^{2}}\oplus
{\cal H}^{n}(\sm,j)d\mu(\sm)
\end{equation}
of the irreducible representations 
\begin{equation}
\label{irrep}
{\cal H}(\sm,j),\qquad 
\sm=p_{\mu}p^{\mu}=E-\bs{p}^{2}\,.
\end{equation}
If we consider in (\ref{directsum}) only 
(continuous) direct sums with the same value
for $j=j_{R}$ then ${\cal U}(\Lambda)$ for any Lorentz transformation
$\Lambda$ is, according to (\ref{wignerrotation}), 
the same operator function of 
the $6$ parameters
which are given by the three $\p^{m}$ or the three $v^{m}$ :
\begin{equation}
\label{uandv}
\left(
\begin{array}{c}
\p^{0} \\
\p^{m}
\end{array}
\right)=\left(
\begin{array}{c}
\left( 1-\frac{\bs{v}^{2}}{c^{2}} \right)^{-\frac{1}{2}} \\
\left( 1-\frac{\bs{v}^{2}}{c^{2}} \right)^{-\frac{1}{2}}v^{m}
\end{array}\right)
\end{equation}
and the three rotation angles (e.g., Euler angles in the rest frame).
The analytic continuation in $\sm$ can therefore be accomplished without
affecting the Lorentz transformations. The Lorentz transformations in 
the minimally-complex mass representation are represented unitarily
by the same operators ${\cal U}(\Lambda)$ as in Wigner's UIR $(m,j_{R})$.
At rest, on $|\bs{0},j_{3}\, (\sm,j_{R})\>$, only the time translations
of $\cal P$ will be represented non-unitarily for complex values of $\sm$.
And using (\ref{boosthat}) only the label $\sm$ in the velocity basis 
$|\bs{\p},j_{3}\, (\sm,j_{R})\>$ is complex. 
The basis vector decomposition (\ref{dirachat2}) using the velocity basis,
\begin{equation}
\label{dirachat3}
\phi= \sum_{j_{3}}\int d\mu(\sm)\int d\mu (\bs{\p})
|\,\bs{\p},j_{3}(\sm,j)\rangle
\langle (\sm,j)j_{3},\bs{\p}\,|\,
\phi\rangle 
\quad\text{for } \phi \in \Phi \subset {\cal H}(\sm,j)\, ,
\end{equation}
is therefore more suitable than (\ref{diracbasisvectors})
that uses the momentum basis,
because $\bs{\p}$ is independent of $\sm$ while
$\bs{p}=\sqrt{\sm}\bs{\p}$ is not. If we deform the contour
of integration for $\sm$ from the real axis as in (\ref{directsum}) into the 
complex $\sm$-plane then the integral over $d\mu(\bs{\p})$ in (\ref{dirachat3})
remains unaffected.

\section{Relativistic Kinematics for (two-particle) Resonance Scattering}

Continuous direct sums like (\ref{directsum}) appear in the case 
of scattering experiments of two relativistic particles like e.g.,
the process
\begin{mathletters}
\label{resonancescattering}
\begin{equation}
\label{twoelectronsresonance}
e^{+}\, e^{-} \rightarrow \rho^{0} \rightarrow
\pi^{+}\, \pi^{-} \, ,
\end{equation}
or the more theoretical process
\begin{equation}
\label{twopionsresonance}
\pi^{+}\,\pi^{-} \rightarrow \rho^{0} \rightarrow
\pi^{+}\, \pi^{-} \, .
\end{equation}
\end{mathletters}
These processes predominantly happen in the $j^{P}=1^{-}$ partial 
amplitude if the
$\rho$-meson mass region is selected for the invariant mass square
\begin{equation}
\label{invariantmass}
\sm =(p_{1}+p_{2})^{2}=E_{\rho}^{2}+\bs {p}_{\rho}^{2}\, ,\,\, 
E_{\rho}=E_{1}+E_{2}\, ,\,\,
\bs{p}_{\rho}=\bs{p_{1}}+\bs{p_{2}}\, ,
\end{equation}
where $p_{1}$ and $p_{2}$ are the momenta
of the two pions $\pi^{+}$,$\pi^{-}$
\footnote{
Though our discussions apply with obvious modifications
to the general case of
$$1+2+3+\cdots \rightarrow R_{i}\rightarrow 1^{'}+2^{'}+3^{'}+\cdots$$
these generalizations lead to enormously more complicated equations.
For the sake of simplicity, we shall therefore 
consider a resonance scattering process like (\ref{resonancescattering}).}.
The relativistic one particle states are given by an irreducible 
representation space 
${\cal H}^{n_{i}}(m_{i},s_{i})$ of the Poincar\'e group $\cal{P}$.
The independent, \emph{interaction-free} 
two-particle states (or $n$ particle states)---
like the $\pi^{+}\, \pi^{-}$ system in (\ref{twopionsresonance})---
are given by the direct product of the irreducible representation
spaces ${\cal H}(m_{1},s_{1})$ and ${\cal H}(m_{2},s_{2})$ :
${\cal H}^{n_{1}}(m_{1},s_{1})
\otimes{\cal H}^{n_{2}}(m_{2},s_{2})\equiv {\cal H}$.
Empirical evidence suggests that the resonances in processes like 
(\ref{resonancescattering}) appear in one partial amplitude
with a given value of resonance spin $j_{R}$ (e.g., $j_{\rho}^{P}=1^{-}$).
Therefore, the first problem is the reduction of the direct product
${\cal H}(m_{1},s_{1})\otimes{\cal H}(m_{2},s_{2})$ into a direct sum
of ${\cal H}^{n}(\sm,j)$; the second problem is how to 
go from the free two-particle system to the interacting two-particle
system. 

The first problem has been solved in general 
\cite{aW60,hJ62,aM62}
\begin{equation}
\label{reduction}
{\cal H}\equiv {\cal H}^{n_{1}}(m_{1},s_{1})\otimes
{\cal H}^{n_{2}}(m_{2},s_{2})
=\int_{(m_{1}+m_{2})^{2}}^{\infty}d\mu(\sm)\sum_{nsl}\sum_{j}
\oplus{\cal H}^{nsl}(\sm,j)\,.
\end{equation}
The sums in (\ref{reduction}) extend over 
$$
j=
\begin{array}{ccccl}
0 & 1 & \cdots & \text{ if } & s_{1}+s_{2}=\text{ integer }\\
1/2 & 3/2 & \ldots & \text{ if } & s_{1}+s_{2}=\text{ half integer }
\end{array}\, ,
$$
and the degeneracy indices $(l,s)$ for a given $j$ are summed over
\begin{eqnarray}
\nonumber
s&=&s_{1}+s_{2}\, ,\, s_{1}+s_{2}-1\, ,\, \ldots |s_{1}-s_{2}|\\
\nonumber l&=&j+s\, ,\, j+s-1\, ,\, j+s-2\, ,\, \ldots j-s \, .
\end{eqnarray}
Here $j$ represents the total angular momentum of the combined
$\pi^{+}\pi^{-}$ system; one of these values will be the resonance spin
$j_{R}$. The degeneracy indices $(s,l)$ for each fixed value of $j$
are the total spin angular momentum and the total orbital angular
momentum of the two $\pi$, respectively.
The quantum number $n$ is summed over all channel numbers that can be obtained
by combining the species numbers $n_{1}$ and $n_{2}$ of the two $\pi$.

Instead of the invariant mass square $\sm=p_{\mu}\,p^{\mu}=E^{2}-
\bs{p}^{2}$ that we have used in 
(\ref{reduction}) one often uses $w=\sqrt{\sm}$, the invariant mass
or the energy in the center of mass system of the two particles $n_{1},
n_{2}$ \cite{aW60,hJ62,aM62}. The choice of the measure 
\begin{equation}
\eqnum{\ref{reduction}$a$}
d\mu(\sm)=\rho(\sm)d\sm,\quad (\text{or if one uses $w$, of } 
d\mu(w)=\rho(w)dw)
\end{equation}
depends upon the normalization of the system
of generalized basis vectors of (\ref{reduction}). We shall use
\begin{equation}
\label{choice}
\eqnum{\ref{reduction}$b$}
\rho(\sm)=1,\text{ and then } \rho(w)=2w 
\end{equation}
if we label the basis by $w$ so that we do not change 
the ``normalization'' of the kets.
The resonance space will be related (but will not be identical)
to a subspace of (\ref{reduction}) with a definite value of
angular momentum $j$ (e.g., $j=j_{3}^{P}=1^{-}$ 
in case of the $\rho$-resonance 
of (\ref{resonancescattering})). This is based on empirical evidence;
 resonances appear in one particular partial amplitude with a particular
value of resonance spin $j=j_{R}$ 
(though it may happen that there are more than one
resonance in the same partial amplitude, but at different
resonance energy $\sm_{R_{1}},\,\sm_{R_{2}},\,\cdots$). 
We will therefore single
out a particular subspace
\begin{equation}
\label{subspace}
{\cal H}^{nls}=\int_{(m_{1}+m_{2})^{2}}^{\infty}
d\sm \oplus {\cal H}^{nls}(\sm,j)
\end{equation}
with definite degeneracy or/and channel quantum numbers
$\eta=ls,$ $n$.

The reduction (\ref{reduction}) is usually done using the Wigner
momentum kets (\ref{wignerbasis}) in which the Clebsch-Gordan
coefficients are given by \cite{aW60,hJ62,aM62} :
\begin{equation}
\label{cg}
\langle\, p_{1}s_{13}\,p_{2}s_{23}\,[m_{1}s_{1}, m_{2}s_{2}]\,|\,
pj_{3}\,[wj],\eta\,\rangle, 
\text{ where $\eta$ now denotes }\eta=n,l,s\,.
\end{equation}
For the reasons mentioned above we want to work
with the 4-velocity eigenkets
$|\hat{p},j_{3}\,[w,j],\eta\,\rangle$
which are eigenvectors of the operators
\begin{equation}
\label{operators}
\hat{P}_{\mu}=({P}^{(1)}_{\mu}+{P}^{(2)}_{\mu})M^{-1},\,\,
M^{2}=(P^{(1)}_{\mu}+P^{(2)}_{\mu})
(P^{(1)\mu}+P^{(2)\mu})
\end{equation}
with eigenvalues
\begin{equation}
\label{eigenvalues}
\hat{p}^{\mu}=
\left(
\begin{array}{c}
\hat{E}=\frac{p^{0}}{w}=\sqrt{1+\bs{\hat{p}}^{2}}=\p^{0}\\
\bs{\hat{p}}=\frac{\bs{p}}{w}
\end{array} \right)
\text{ and eigenvalues }
w^{2}=\sm\, .
\end{equation}
In here $\hat{P}^{(i)}_{\mu}$ are the 4-velocity operators in the one
particle spaces ${\cal H}^{n_{i}}(m_{i},s_{i})$ with eigenvalues
$\hat{p}^{i}_{\mu}=\frac{p^{i}_{\mu}}{m_{i}}$.
The Clebsch-Gordan coefficients are the transition coefficients
$\langle\hat{p}_{1}\hat{p}_{2}\,s_{13}s_{23}\,[m_{1}s_{1},m_{2}s_{2}]\,
|\,\hat{p}j_{3}\,[wj],\eta\,\rangle$
between the direct product basis
\begin{equation}
\label{directproductbasis}
|\hat{p}_{1}s_{13}\,m_{1}s_{1}\rangle\otimes
|\hat{p}_{2}s_{23}\,m_{2}s_{2}\rangle
\equiv
|\hat{p}_{1}\hat{p}_{2}\,s_{13}s_{23}\,[m_{1}s_{1},m_{2}s_{2}]\,\rangle
\end{equation}
and the angular momentum basis 
$|\p j_{3}\,[wj],\eta\,\>$.

To obtain the Clebsch-Gordan coefficients, one follows the same
procedure as given in the classic papers \cite{aW60,hJ62,aM62} 
for the Clebsch-Gordan coefficients
(\ref{cg}). This will be done in the Appendix, where the general
case will be discussed. Here we shall restrict 
ourselves to the special case $s_{1}=0,s_{2}=0$ to avoid the
inessential complications due to the $SO(3)$ Clebsch-Gordan coefficients
for the angular momentum couplings $s_{1}\otimes s_{2}\rightarrow s$,
$s\otimes l \rightarrow j$ and the occurrence of the Wigner
rotations $R(L^{-1}(\hat{p}),\p_{i})$ of the inverse
boost $L^{-1}(\hat{p})$ which will enter in (\ref{cg}).
Also for the process (\ref{twopionsresonance}) this is sufficient,
since $s_{{\pi}^{+}}=s_{{\pi}^{-}}=0$. There is no degeneracy
of the angular momentum basis vectors in this case and 
$|\p j_{3}\,[wj]\,\>$ is given in terms of (\ref{directproductbasis})
by
\begin{eqnarray}
\label{expansionhat}
&|\,\hat{p}j_{3}\,[wj]\,\rangle=\int\frac{d^{3}\hat{p}_{1}}{2\hat{E}_{1}}
\frac{d^{3}\hat{p}_{2}}{2\hat{E}_{2}}
\,|\,\hat{p}_{1}\hat{p}_{2}[m_{1}m_{2}]\,\rangle\,\langle\,
\hat{p}_{1}\hat{p}_{2}[m_{1}m_{2}]\,|\,\hat{p}j_{3}\,[wj]\,\rangle\\
\nonumber
&\text{for any }(m_{1}+m_{2})^{2}\leq w^{2}<\infty\qquad j=0,1,\cdots
\end{eqnarray}
The choice of the measure $\frac{d^{3}\hat{p}_{i}}{2\hat{E}_{i}(\bs{\p}_{i})}
=\frac{d^{3}p_{i}}{m_{i}^{2}2E_{i}}$ is the same as
(\ref{dirachat}).

From the 4-translation invariance (conservation of
$4$-momentum) it follows that the Clebsch-Gordan is 
of the form
\begin{equation}
\label{deltafour}
\langle\,\hat{p}_{1}\hat{p}_{2}\,|\,\hat{p}j_{3}\,[wj]\,\rangle
=\delta^{4}(p-r)\langle\!\langle\,\hat{p}_{1}\hat{p}_{2}\,|\,\hat{p}j_{3}
\,[wj]\,\rangle\!\rangle\, ,\quad \text{where } r\equiv p_{1}+p_{2}\, .
\end{equation}
The reduced matrix element in the center-of-mass is in analogy to the 
non-relativistic case given by \cite{aB93}
\begin{equation}
\label{reducedelement}
\langle\!\langle\,\hat{p}_{1}^{cm}\hat{p}_{2}^{cm}\,|\,
\bs{0}j_{3}\,[wj]\,\rangle\!\rangle=Y_{jj_{3}}(\bs{e})
\tilde\mu_{j}(w,m_{1},m_{2})\, ,
\end{equation}
where $\tilde{\mu}_{j}(w,m_{1},m_{2})$ 
is a function of $w$ (or $\sm$) which depends
upon our choice of ``normalization'' for the basis vectors 
$|\p j_{3}\,[wj]\,\>$ in (\ref{expansionhat}).
The equations (\ref{deltafour}) and (\ref{reducedelement})
are combined into 
\begin{eqnarray}
\label{cgseries}
&\langle\,\hat{p}_{1}\hat{p}_{2}\,|\,\hat{p}j_{3}\,[wj]\,\rangle
=2\hat{E}(\bs{\p})\delta^{3}(\bs{p}-\bs{r})
\delta(w-\epsilon)Y_{jj_{3}}(\bs{e})\mu_{j}(w,m_{1},
m_{2})\\
\nonumber &\text{ with }\epsilon^{2}=r^{2}=(p_{1}+p_{2})^{2} \, ,
\end{eqnarray}
where again $\mu_{j}(w,m_{1},m_{2})$ is a function
that fixes the $\delta$-function ``normalization'' of 
$|\p j_{3}\,[wj]\,\>$.
The unit vector $\bs{e}$ in (\ref{reducedelement}) 
is chosen to be in the c.m.\ frame
the direction of $\bs{\hat{p}}_{1}^{cm}=-\frac{m_{2}}{m_{1}}\,\bs{\hat{p}}_{2}
^{cm}$. In general it is obtained from the relative ``4-momentum''
$q_{\mu}$ of Michel and Wightman \cite{aW60} 
by $e_{i}=L^{-1}(p)^{.\,\mu}_{i}q_{\mu}$. 
The $\mu_{j}(w,m_{1},m_{2})$ and $\tilde\mu_{j}(w,m_{1},m_{2})$ 
are some weight functions
which are determined from the required ``normalization'' of the
4-velocity kets (\ref{expansionhat}). Since for a fixed value
of $[wj]$ these generalized eigenvectors are the basis of the
irreducible representation space ${\cal H}(w,j)$ of the Poincar\'e group,
we want them to be normalized like (\ref{measurehat}),
which in (\ref{expansionhat}) has been already
assured by the choice of the invariant measure
$\frac{d^{3}\p_{i}}{2\hat{E}_{i}}$.
Therefore, in analogy to (\ref{normalizationhat}), 
we take for the normalization
of the basis vectors (\ref{expansionhat}) to be
\begin{eqnarray}
\label{normalizationchoice}
\langle\,\hat{p}'j_{3}'\,[w'j']\,|\,\hat{p}j_{3}\,[wj]\,\rangle\,
=2\hat{E}(\bs{\p})\delta^{3}(\bs{\hat{p}'}
-\bs{\hat{p}})\delta_{j_{3}'j_{3}}\delta_{j'j}\delta(\sm-\sm')\, ,\\
\nonumber
\text{ where }\hat{E}(\bs{\p})=\sqrt{1+\bs{\p}^{2}}
=\frac{1}{w}\sqrt{w^{2}+\bs{p}^{2}}\equiv\frac{1}{w}E(\bs{p},w)\, .
\end{eqnarray}
The $\delta$-function normalization $\delta(\sm'-\sm)=\frac{1}
{2w}\delta(w-w')$ in (\ref{normalizationchoice}) 
is a consequence of the choice (\ref{choice}) for the measure.
After we have chosen the normalization as in 
(\ref{normalizationchoice}), one determines
the weight function $\mu_{j}(w,m_{1},m_{2})$
using (\ref{expansionhat}). The result is :
\begin{equation}
\label{weight}
\left| \mu_{j}(w,m_{1},m_{2})
\right|^{2}=\frac{2m_{1}^{2}m_{2}^{2}w^{2}}
{\sqrt{\lambda(1,(\frac{m_{1}}{w})^{2},(\frac{m_{2}}{w})^{2})}}\, ,
\end{equation}
where $\lambda$ is defined by \cite{aW60}:
$$
\lambda(a,b,c)=a^{2}+b^{2}+c^{2}-2(ab+bc+ac)\,.
$$
Except for the normalization factor $\mu$, which follows from
our chosen normalization (\ref{normalizationchoice}), 
the values of the Clebsch-Gordan 
coefficients (\ref{cgseries}) is quite obvious 
\footnote{
A formula like (\ref{cgseries}) is also given and explained in section
$3.7$ of \cite{sW95} which for $s=0,\, s_{1}=s_{2}=0$
agrees with (\ref{cgseries}) except for the normalization factor
(\ref{weight}). For $s\ne 0,\, s_{i}\ne 0$, see Appendix.}.
It expresses momentum conservation and the only factor
that one may be puzzled about is that it should be consistent with
the 4-velocity normalization expressed by the
$\delta^{3}(\bs{\hat{p}'}-\bs{\hat{p}})$,
$\bs{\hat{p}}=\frac{\bs{p}}{w}$ in 
(\ref{normalizationchoice}).
Therewith, we have obtained by (\ref{expansionhat}) with 
(\ref{directproductbasis}) and (\ref{cgseries})
a system of basis vectors for the space (\ref{reduction})
(with $s_{1}=s_{2}=0$) which is the representation space
of scattering processes like (\ref{twopionsresonance}).
As expected, the basis vectors are outside the Hilbert space;
$|\,\hat{p}j_{3}\,[wj]\,\rangle\in\Phi^{\times}\supset{\cal H}\supset\Phi$.
They have definite values of angular momentum $j$ and invariant
mass $w\equiv\sqrt{\sm}$
\footnote{
Written in terms of Hilbert spaces, $d\mu(\sm)$ means Lebesgue integrations.
However, within the RHS mathematics, one can choose for
$\langle\, \phi\,|\,\hat{p}j_{3}\,[wj]\,\rangle$ a smooth function
and use Riemann integration and assign to each vector a well defined value
$w$ (not just up to a set of measure zero)}; we shall define the Gamow vectors
(describing $\rho^{0}$) in terms of linear combinations of these
c.m.-energy eigenvectors with a definite value of $j$. However,
since the resonances form and decay under the influence of an
interaction and the $|\hat{p}j_{3}\,[wj]\,\rangle$ are interaction-free
eigenvectors of the ``free-particle'' Hamiltonian
\begin{equation}
\label{freehamiltonian}
K=P_{1}^{0}+P_{2}^{0}
\end{equation}
we have to go from the free-particle basis vectors 
(\ref{expansionhat}) to the interaction-basis vectors.
This can be done in analogy to the non-relativistic case and may be 
justified in two ways :\\
1) One assumes that the time translation generator for the interaction
system has two terms (\cite{sW95} Ch.3), 
$H_{0}$ and the interaction $V$
\begin{equation}
\label{interactionhamiltonian}
H=H_{0}+V
\end{equation}
in such a way that to each eigenvector of $H_{0}$ with
eigenvalue $E=w\sqrt{1+{\bs{\p}}^{2}}$,
\begin{equation}
\label{freeeigenvalue}
H_{0}\,|\,\hat{p}j_{3}\,[wj]\,\rangle=E\,|\,\hat{p}j_{3}\,[wj]\,\rangle\,,
\end{equation}
there correspond eigenvectors of $H$ with the same eigenvalue
\begin{equation}
\label{interactioneigenvalue}
H|\hat{p}j_{3}\,[wj]^{\pm\, int}\,\rangle=E\,|\,\hat{p}j_{3}\,[wj]^{\pm\, int }
\,\rangle \, .
\end{equation}
Since vectors are not completely defined by the requirement that
they be eigenvectors of an operator with a given eigenvalue
(but may differ by a phase factor (phase shifts) or unitary transformation
(S-matrix) in case of degeneracy) we have added the additional label int.
This additional specification of the eigenvectors can be chosen in a 
variety of ways that are connected with the spaces $\Phi$ that one admits,
i.e., with initial and final boundary conditions (as explained for the 
non-relativistic case in \cite{aB93}).
Since (\ref{interactionhamiltonian}) may be a questionable 
hypothesis in relativistic physics a second justification does not
make use of the existence of the Hamiltonian splitting 
(\ref{interactionhamiltonian}).\\
2) One assumes the existence of an $S$-operator and of 
M{\o}ller operators $\Omega^{+}$ and $\Omega^{-}$. 
$\Omega^{+}$ transforms non-interacting states $\phi^{in}$ which
are prepared by an apparatus far away from the interaction region
into exact state vectors $\phi^{+}$, 
\begin{equation}
\label{freein}
\Omega^{+}\phi^{in}=\phi^{+}\, ,\quad
\phi^{+}(t)=e^{-iHt}\phi^{+}\, ,
\end{equation}
which evolve with the exact time-evolution operator $H$.
$\Omega^{-}$ transforms observables $|\psi^{out}\rangle\langle\psi
^{out}|$ registered by the detector placed far away from the 
interaction region into the vectors $\psi^{-}$ which evolve with
the exact $H$ in the interaction region :
\begin{equation}
\label{freeout}
\Omega^{-}\psi^{out}=\psi^{-}\, ,
\quad
\psi^{-}(t)=e^{iHt}\psi^{+}\, ,
\end{equation}
where $t$ is the time in the c.m.\ frame.
The basis vectors for the free-particle space and the interaction-basis 
vectors are then assumed to be related by
\footnote{In non-relativistic scattering off a fixed target one assumes 
that the $|\bs{p}^{+}\rangle$ related by (\ref{moeller})
to the $|\bs{p}\rangle$ are not eigenvectors
of $\bs{P}$ since $[V,P_{i}]\ne 0$.}
\begin{equation}
\label{moeller}
|\,\hat{p}j_{3}\,[wj]^{\pm}\,\rangle=\Omega^{\pm}|\,\hat{p}j_{3}\,[wj]\,\rangle
\, .
\end{equation}
If (\ref{interactionhamiltonian}) also holds then the symbol 
$\Omega^{\pm}$ at the center-of-mass is given by the solution of the
Lippmann-Schwinger equation
\begin{equation}
\label{interactioninout}
|\bs{0}j_{3}\,[wj]^{\pm}\,\rangle=\left(1+\frac{1}{w-H\pm i\epsilon}V\right)
|\bs{0}j_{3}\,[wj]\,\rangle\, .
\end{equation}
The vectors $|\hat{p}j_{3}\,[wj]^{\pm}\,\rangle$ are obtained from the basis
vectors at rest $|\bs{0}j_{3}\,[wj]^{\pm}\,\rangle$ by the boost (rotation-free
Lorentz transformation) ${\cal U}(L(\hat{p}))$ whose parameters
are the $\hat{p}^{m}$ and whose generators are the 
interaction-incorporating observables
\begin{equation}
\label{interactionobservables}
P_{0}=H,\quad P^{m},\quad J_{\mu\nu} \, ,
\end{equation}
i.e., the exact generators of the Poincar\'e group (\cite{sW95} section $3.3$).
These vectors (\ref{interactioninout}), which for a fixed value of 
$[wj]$ span an irreducible representation space of the Poincar\'e 
group with the ``exact generators'', will be used for the definition
of the relativistic Gamow vectors. The values
of $j$ and $\sm=w^{2}$ are $j=\text{ integer }$
(for $s_{1}=s_{2}=0$ otherwise also half integer) and
$(m_{1}+m_{2})^{2}\leq \sm <\infty$.
The value of $j$ will be fixed and represents the resonance spin;
the same we do with parity and the degeneracy quantum numbers
($n,\,\eta$). The values of $\sm$ we shall continue from the 
physical values into the complex plane of the relativistic
$S$-matrix. 

\acknowledgments

We are grateful for some helpful correspondence with L.~Michel.
Support from the Welch Foundation is gratefully acknowledged.

\appendix
\section*{Reduction of the Direct Product of Two One-Particle UIR
of $\cal P$$\,^8$}

\addtocounter{footnote}{1}
\footnotetext{The discussion here follows the one in \cite{aM62}
with the difference that here the two particle irreducible
representation spaces are labeled by the square of the total invariant mass
$\sm$ instead of $w=\sqrt{\sm}$; and the velocity basis
are used instead of the momentum basis.}
We discuss here the reduction of the direct product of two 
one-particle irreducible representation 
spaces of the Poincar\'e group $[m_{1},s_{1}]\otimes[m_{2},s_{2}]$
into a continuous direct sum of 
irreducible representation (irrep) spaces $[\sm,j]$ of invariant
mass squared $\sm$ and spin $j$. This has been done
in \cite{aW60,hJ62,aM62} using the Wigner
basis systems of momentum eigenvectors. Here we shall do
it using the $4$-velocity basis vectors of the Poincar\'e
group $\cal P$ and obtain the Clebsch-Gordan coefficients of 
$\cal P$ for the velocity basis.
For the one particle spaces, we choose the c.s.c.o.\ (\ref{cscohat})
with the generalized eigenvectors (\ref{wignerbasishat}).
Thus, the one particle spaces $\H (m,j)$ are labeled by the
mass $m$ and the spin $j$ 
of the particle. In analogy to the case of one-particle,
a two-particle irrep space is labeled by the square
of the total invariant mass 
$\sm=(p_{1}+p_{2})^{2}$ and the total angular momentum 
$j$ of the two particles. The two-particle irrep space is
denoted by $\H^{\eta}_{n}(\sm,j)$, where $\eta$ is a degeneracy
label and $n$ is a particle species label. 
Thus the reduction problem is written as
\begin{equation}
\label{a:reduction}
\H(m_{1},s_{1})\otimes \H(m_{2},s_{2})
=\sum_{j\eta}\int_{(m_{1}+m_{2})^{2}}^{\infty}
\oplus \H^{\eta}_{n}(\sm,j) d\sm \, .
\end{equation}
As in (\ref{wignerbasishat}), the two-particle basis vectors of 
$\H^{\eta}_{n}(\sm,j)$ have as the only continuous variables
the total four velocity of the two particles and
the square of the total invariant mass of the two particles.
These basis vectors are denoted by :
\begin{equation}
\label{a:basishat}
|\p\s[\sm j]\eta,n\>
\end{equation}
with the normalization :
\begin{equation}
\label{a:normalizationhat}
\<\,\p'\s'[\sm'j']\eta',n'\,|\,\p\s[\sm j]\eta,n\,\>
=2\p_{0}\,\delta_{nn'}\delta_{jj'}\delta_{\s\s'}
\delta_{\eta\eta'}
\delta^{3}(\bs{\p-\p'})\delta(\sm-\sm')\,,
\end{equation}
where $\s$ is the three-component of the total angular
momentum $j$.
We denote the basis vectors of $\H(m_{1},s_{1})\otimes
\H(m_{2},s_{2})$ by :
\begin{equation}
\label{a:directproductbasis}
|\,\p_{1}\s_{1}[m_{1}s_{1}]\,\>\otimes 
|\,\p_{2}\s_{2}[m_{2}s_{2}]\,\>\equiv
|\,\p_{1}\s_{1}[m_{1}s_{1}],\p_{2}\s_{2}[m_{2}s_{2}]\,\>\,,
\end{equation}
where $\s_{1}\, ,\s_{2}$ are the three-components of the 
spins $s_{1}\, ,s_{2}$ respectively.
In order to obtain the Clebsch-Gordan coefficients,
\begin{equation}
\label{a:clebschgordon}
\<\,\p_{1}\s_{1}[m_{1}s_{1}],\p_{2}\s_{2}[m_{2}s_{2}]\,|\,
\p \s [\sm j]\eta, n\,\>\, ,
\end{equation}
of the reduction (\ref{a:reduction}), 
we start by relabeling the basis vectors in 
(\ref{a:directproductbasis}) by using $\sm$, $\p$ and the unit vector 
$\bs{\hat{n}}=\frac{\bs{p_{1}}-\bs{p_{2}}}{|\bs{p_{1}}-\bs{p_{2}}|}$
as continuous parameters (we note that both sets,
$\{\p_{1},\, \p_{2}\}$ and $\{\p,\, \bs{\hat{n}},\,\sm\}$  consist
of six independent parameters).
Thus, we can write :
\begin{equation}
\label{a:relabel}
|\,\p_{1}\s_{1}[m_{1}s_{1}],\,\p_{2}\s_{2}[m_{2}s_{2}]\,\>
\equiv
|\,\p\,\bs{\hat{n}}\,\sm,\,\s_{1}[m_{1}s_{1}]\s_{2}[m_{2}s_{2}]\,\>\,.
\end{equation}
In the rest frame of both particles, i.e., for 
$\p=
\p_{R}=\scriptstyle{\left(
\begin{array}{cccc}
1 \\ 0 \\0 \\0
\end{array}
\right)}\, , $ we can expand the unit vector $\bs{\hat{n}}$ 
in terms of orbital
angular momentum basis vectors :
\begin{equation}
\label{a:sh}
|\bs{\hat{n}}\>=\sum_{ll_{3}}|ll_{3}\>\<ll_{3}|\bs{\hat{n}}\>
=\sum_{ll_{3}}|ll_{3}\>Y^{*}_{ll_{3}}(\bs{\hat{n}})\, .
\end{equation}
We can further use the angular momentum Clebsch-Gordan coefficients
to combine the two spins, $s_{1}$ and $s_{2}$, to give a total spin
$s$ with three component $\mu$, which in turn is added to the
orbital angular momentum $l$ with three component $l_{3}$ to form
a total angular momentum $j$ with three component $\s$.
This gives the basis vector for the two-particle irrep space
\begin{equation}
\label{a:basistwoparticles}
|\,\p\s[\sm j]ls;\,m_{1}s_{1},m_{2}s_{2}\>\, .
\end{equation}
Thus, the degeneracy label $\eta$ in (\ref{a:basishat}) 
designates the total spin $s$ 
and the total orbital angular momentum $l$ of both particles;
and the masses 
$m_{1},\,m_{2}$ and spins $s_{1},\,s_{2}$  
of both particles are included in the particle species label $n$.
Thus, (\ref{a:reduction}) can be rewritten in more
details as :
\begin{equation}
\label{a:reductiondetails}
\H(m_{1},s_{1})\oplus \H(m_{2},s_{2})
=\sum_{jls}\int_{(m_{1}+m_{2})^{2}}^{\infty}
\oplus \H_{n}^{ls}(\sm,j)d\sm\, ,
\end{equation}
\begin{eqnarray}
\nonumber
\text{where  }\quad s&=&|s_{1}-s_{2}|,\,|s_{1}-s_{2}|+1,\,
\cdots\, ,s_{1}+s_{2}\\
\nonumber 
j &=& |l-s|,\,|l-s|+1,\,\cdots\, ,l+s \, .
\end{eqnarray}
With (\ref{a:relabel}) and (\ref{a:sh}), we deduce that in the 
rest frame, the Clebsch-Gordan coefficients of (\ref{a:reductiondetails})
are given by :
\begin{eqnarray}
\nonumber
\<\,\p_{1}\s_{1}[m_{1}s_{1}],\p_{2}\s_{2}[m_{2}s_{2}],n\,|\,
\p_{R}\s[\sm j],\eta,n'\,\> 
 &=& 2N_{n}(\sm) \delta_{nn'}
\theta(\sm-(m_{1}+m_{2})^{2})
\delta^{3}(\bs{p_{1}+p_{2}})
\delta(\sm-(p_{1}+p_{2})^{2})\\ \label{a:cgrest}
&& \times \sum_{l_{3}\mu}C_{s_{1}s_{2}}(s\mu,\s_{1}\s_{2})
C_{sl}(j\s,\mu l_{3})Y_{ll_{3}}(\bs{\hat{n}})\, ,
\end{eqnarray}
where $N_{n}(\sm)$ is a normalization factor.
Having obtained the Clebsch-Gordan coefficients in the rest frame
(\ref{a:cgrest}), we can use the boost operator (\ref{boosthat}) 
to obtain the Clebsch-Gordan coefficients in a general frame 
\footnote{Formula $(3.7.5)$ in \cite{sW95}, which corresponds
to (\ref{a:cg}) but for different choices of basis and normalizations,
is missing the rotation matrices factors that appear in the 
Clebsch-Gordan coefficients away from the rest frame, as exhibited
in (\ref{a:cg}).}:
\begin{eqnarray}
\nonumber
\<\,\p_{1}\s_{1}[m_{1}s_{1}],\p_{2}\s_{2}[m_{2}s_{2}],n\,|\,
\p\s [\sm j]\eta,n'\,\>
&=& \<\,\p_{1}\s_{1}[m_{1}s_{1}],\p_{2}\s_{2}[m_{2}s_{2}],n\,|\, 
{\cal U}(L(p))\,|\,\p_{R}\s [\sm j]\eta,n'\,\> \\
\nonumber
&=&2\p_{0}N_{n}(\sm)\delta_{nn'}\theta(\sm-(m_{1}+m_{2})^{2})
\delta^{3}(\bs{p-p_{1}-p_{2}})\\ 
\nonumber
&& \times \delta(\sm-(p_{1}+p_{2})^{2})
\sum_{\s_{1}'\s_{2}'}D^{s_{1}*}_{\s_{1}'\s_{1}}(
R(L^{-1}(p),p_{1}))D^{s_{2}*}_{\s_{2}'\s_{2}}(
R(L^{-1}(p),p_{2}))\\ \label{a:cg}
&&\times \sum_{l_{3}\mu}C_{s_{1}s_{2}}(s\mu,\s_{1}'\s_{2}')
C_{sl}(j\s,\mu l_{3})\, Y_{ll_{3}}(\bs{e})\, ,
\end{eqnarray}
where $R(\lambda,p)$ is the Wigner rotation given in (\ref{wignerrotation})
and 
$$
\bs{e}=\frac{\overrightarrow{L^{-1}(p)(p_{1}-p_{2})}}
{\left| \overrightarrow{L^{-1}(p)(p_{1}-p_{2})} \right| }
\, .
$$
The normalization factor $N_{n}(\sm)$ depends upon our 
normalization choice (\ref{a:normalizationhat}). 
Before discussing how to obtain it,
let us first introduce the following notations :
\begin{mathletters}
\label{a:notations}
\begin{eqnarray}
&\Gamma(s_{1}\s_{1},s_{2}\s_{2},s\mu)
=\sum_{\s_{1}'\s_{2}'}
D^{s_{1}*}_{\s_{1}'\s_{1}}(R(L^{-1}(p),p_{1})) 
D^{s_{2}*}_{\s_{2}'\s_{2}}(R(L^{-1}(p),p_{2}))
C_{s_{1}s_{2}}(s\mu,\s_{1}'\s_{2}')\, ,\\
&Y_{j\s ls}(\bs{e},\mu)=\sum_{l_{3}}C_{sl}(j\s,\mu l_{3})
Y_{ll_{3}}(\bs{e})\, .
\end{eqnarray}
\end{mathletters}
With the above notations, (\ref{a:cg}) is written as 
\begin{eqnarray}
\nonumber
\<\,\p_{1}\s_{1}[m_{1}s_{1}],\p_{2}\s_{2}[m_{2}s_{2}],n\,|\,
\p\s[\sm j]\eta,\, n'\>
&=& 2\p_{0}N_{n}(\sm)\,\delta_{nn'}\delta^{3}(\bs{p-p_{1}-p_{2}})
\delta(\sm-(p_{1}+p_{2})^{2})\\ \label{a:cg2}
&& \times\sum_{\mu}\Gamma(s_{1}\s_{1},s_{2}\s_{2},s\mu)
Y_{j\s ls}(\bs{e},\mu)
\,.
\end{eqnarray}
In order to obtain the normalization factor $N_{n}(\sm)$,
we insert a complete set of basis vectors (\ref{a:directproductbasis}) in 
$\<\,\p'\s'[\sm'j']\eta',n'\,|\,\p \s[\sm j]\eta,n\,\>$ and use
(\ref{a:cg2}). Upon doing so, we obtain :
\begin{eqnarray}
\nonumber
\<\,\p'\s'[\sm'j']\eta',n'\,|\,\p \s [\sm j]\eta,n\,\> 
&=&\sum_{n''\s_{1}\s_{2}}
\int\frac{d^{3}\p_{1}}{2\p_{1}^{0}}\frac{d^{3}\p_{2}}{2\p_{2}^{0}}
\<\,\p'\s'[\sm'j']\eta',n'\,|\,\p_{1}\s_{1}[m_{1}s_{1}],
\p_{2}\s_{2}[m_{2}s_{2}],n''\,\>\\
\nonumber
&&\times \<\,\p_{1}\s_{1}[m_{1}s_{1}],\p_{2}\s_{2}[m_{2}s_{2}],n''\,|
\,\p \s [\sm j]\eta,n\,\>\\
\nonumber
&=& (2\p_{0})^{2}|N_{n}(\sm)|^{2} \delta_{nn'}
\delta^{3}(\bs{p-p'}) \delta(\sm-\sm')\\
\nonumber
&& \times \sum_{\s_{1}\s_{2}}\sum_{\mu\mu'}\int 
\frac{d^{3}\p_{1}}{2\p_{1}^{0}}
\frac{d^{3}\p_{2}}{2\p_{2}^{0}}
\delta^{3}(\bs{p-p_{1}-p_{2}})\delta(\sm-(p_{1}+p_{2})^{2})\\
\label{a:cg3}
&& \times \Gamma^{*}(s_{1}\s_{1},s_{2}\s_{2},s'\mu')
\Gamma(s_{1}\s_{1},s_{2}\s_{2},s\mu)
Y^{*}_{j'\s'\eta'}(\bs{e},\mu')Y_{j\s\eta}(\bs{e},\mu)\, .
\end{eqnarray}
Using the unitarity of the rotation matrices :
$$
\sum_{\s}D^{*j}_{\s'\s}D^{j}_{\s''\s}=\delta_{\s'\s''}
$$
and the identity
$$
\sum_{\s_{1}\s_{2}}C_{s_{1}s_{2}}(s\mu,\s_{1}\s_{2})
C_{s_{1}s_{2}}(s'\mu',\s_{1}\s_{2})=\delta_{ss'}\delta_{\mu\mu'}\, ,
$$
we find that
\begin{equation}
\label{a:gammaidentity}
\sum_{\s_{1}\s_{2}}\Gamma^{*}(s_{1}\s_{1},s_{2}\s_{2},s'\mu')
\Gamma(s_{1}\s_{1},s_{2}\s_{2},s\mu)
=\delta_{ss'}\delta_{\mu\mu'}\, .
\end{equation}
With the identity (\ref{a:gammaidentity}), (\ref{a:cg3})
can be written as :
\begin{eqnarray}
\nonumber
\<\,\p'\s'[\sm'j']\eta',n'\,|\,p\s[\sm j]\eta,n\,\> &=&
(2\p_{0})^{2}|N_{n}(\sm)|^{2}\delta_{nn'}
\delta^{3}(\bs{p-p'})
\delta(\sm-\sm')\delta_{ss'}
\sum_{\mu l_{3}l_{3}'}C_{sl'}(j'\s',\mu l_{3}')
C_{sl}(j\s,\mu l_{3})\\ \label{a:cg4}
&& \times \int 
\frac{d^{3}\p_{1}}{2\p_{1}^{0}}
\frac{d^{3}\p_{2}}{2\p_{2}^{0}}
\delta^{3}(\bs{p-p_{1}-p_{2}})\delta(\sm-(p_{1}+p_{2})^{2})
Y^{*}_{l'l_{3}'}(\bs{e})Y_{ll_{3}}(\bs{e})\,.
\end{eqnarray}
In order to solve the integration in (\ref{a:cg4}), namely 
\begin{eqnarray}
\nonumber
I &=& \int
\frac{d^{3}\p_{1}}{2\p_{1}^{0}}
\frac{d^{3}\p_{2}}{2\p_{2}^{0}}
\delta^{3}(\bs{p-p_{1}-p_{2}})\delta(\sm-(p_{1}+p_{2})^{2})
Y^{*}_{l'l_{3}'}(\bs{e})Y_{ll_{3}}(\bs{e})\\
\label{a:integration}
&=& \frac{1}{m_{1}^{2}m_{2}^{2}}\int
\frac{d^{3}p_{1}}{2p_{1}^{0}}
\frac{d^{3}p_{2}}{2p_{1}^{0}}
\delta^{3}(\bs{p-p_{1}-p_{2}})\delta(\sm-(p_{1}+p_{2})^{2})
Y^{*}_{l'l_{3}'}(\bs{e})Y_{ll_{3}}(\bs{e})\, ,
\end{eqnarray}
we perform the change of variables (as in equation $(4.9)$ in \cite{aW60}) :
\begin{eqnarray}
\nonumber
p_{1}=\frac{(\sm+m_{1}^{2}-m_{2}^{2})}{2\sm}r+
\frac{\lambda^{1/2}(\sm,m_{1}^{2},m_{2}^{2})}{2\sqrt{\sm}}q \\
\label{a:changeofvariables}
p_{2}=\frac{(\sm-m_{1}^{2}+m_{2}^{2})}{2\sm}r-
\frac{\lambda^{1/2}(\sm,m_{1}^{2},m_{2}^{2})}{2\sqrt{\sm}}q 
\end{eqnarray}
where
$$
\lambda(a,b,c)=a^{2}+b^{2}+c^{2}-2(ab+ac+bc)\, .
$$
With these new variables, we find that
\begin{mathletters}
\label{a:newvariables}
\begin{equation}
\delta(p_{1}^{2}-m_{1}^{2})\delta(p_{2}^{2}-m_{2}^{2})
\delta^{3}(\bs{p}-\bs{p_{1}}-\bs{p_{2}})\delta(\sm-(p_{1}+p_{2})^{2})
=\frac{4\sm^{3/2}}{\lambda^{3/2}(\sm,m_{1}^{2},m_{2}^{2})}\frac{1}{2p_{0}}
\delta(q^{2}+1)\delta(r.q)\delta^{4}(r-p)
\end{equation}
\begin{equation}
d^{4}p_{1}d^{4}p_{2}=\frac{\lambda^{2}(\sm,m_{1}^{2},m_{2}^{2})}{16\sm^{2}}
d^{4}rd^{4}q
\end{equation}
and
\begin{equation}
\bs{e}=\overrightarrow{L^{-1}(p)q}\, .
\end{equation}
\end{mathletters}
Using (\ref{a:newvariables}), the integration (\ref{a:integration})
becomes :
\begin{equation}
\label{a:integration1}
I=\frac{1}{m_{1}^{2}m_{2}^{2}}\frac{1}{2p_{0}}
\frac{\lambda^{1/2}(\sm,m_{1}^{2},m_{2}^{2})}{4\sqrt{\sm}}
\int d^{4}q\delta(q^{2}+1)\delta(p.q)Y_{l'l_{3}'}^{*}\left(\overrightarrow
{L^{-1}(p)q}\right)
Y_{ll_{3}}\left(\overrightarrow{L^{-1}(p)q}\right)\, .
\end{equation}
Performing the change of variable $e=L^{-1}(p)q$ in (\ref{a:integration1}),
we obtain :
\begin{eqnarray}    
\nonumber
I &=& \frac{1}{m_{1}^{2}m_{2}^{2}}\frac{1}{2p_{0}}
\frac{\lambda^{1/2}(\sm,m_{1}^{2},m_{2}^{2})}{8\sm}
\int d\Omega(\bs{e})Y_{l'l_{3}'}^{*}(\bs{e})Y_{ll_{3}}(\bs{e})\\
\label{a:integration2}
&=& \frac{1}{m_{1}^{2}m_{2}^{2}}\frac{1}{2p_{0}}
\frac{\lambda^{1/2}(\sm,m_{1}^{2},m_{2}^{2})}{8\sm}
\delta_{ll'}\delta_{l_{3}l_{3}'}\,.
\end{eqnarray}
Using (\ref{a:integration2}) and the identity
$$
\sum_{\mu l_{3}}C_{sl}(j'\s',\mu l_{3})
C_{sl}(j\s,\mu l_{3})=\delta_{jj'}\delta_{\s\s'}\, ,
$$
(\ref{a:cg4}) finally becomes :
\begin{equation}
\label{a:cg5}
\<\,\p'\s'[\sm'j']\eta',n'\,|\,\p\s[\sm j]\eta,n\,\>
=(2\p_{0})|N_{n}(\sm)|^{2}\frac{1}{m_{1}^{2}m_{2}^{2}}
\frac{\lambda^{1/2}(\sm,m_{1}^{2},m_{2}^{2})}{8\sm^{3}}
\delta_{nn'}\delta_{jj'}\delta_{\s\s'}\delta_{\eta\eta'}
\delta^{3}(\bs{\p-\p'})\delta(\sm-\sm')\, .
\end{equation}
Comparing (\ref{a:cg5}) with (\ref{a:normalizationhat}),
we find that :
\begin{equation}
\nonumber
|N_{n}(\sm)|^{2}=\frac{8m_{1}^{2}m_{2}^{2}\sm^{3}}
{\lambda^{1/2}(\sm,m_{1}^{2},m_{2}^{2})}\, .
\end{equation}

\end{document}